\documentclass[prl,a4paper,twocolumn]{revtex4}

\usepackage{amssymb}
\usepackage{graphicx}
\usepackage{graphics}
\usepackage{amsmath}
\usepackage{amsbsy}
\usepackage{amsthm}
\usepackage{bbm}
\usepackage{bm}
\usepackage{epsfig}

\newcommand{\tr}{\textnormal{Tr}}

\newcommand{\be}{\begin{equation}}
\newcommand{\ee}{\end{equation}}
\newcommand{\beq}{\begin{eqnarray}}
\newcommand{\eeq}{\end{eqnarray}}
\newcommand{\bra}[1]{\ensuremath{\langle #1 |}}
\newcommand{\ket}[1]{\ensuremath{| #1 \rangle}}
\begin{document}

\title{Macroscopic Observables Detecting Genuine Multipartite Entanglement and Partial Inseparability in Many Body Systems}

\author{Andreas Gabriel$^{1}$}
\email{andreas.gabriel@univie.ac.at}
\author{Beatrix Hiesmayr$^{1,2}$}
\email{beatrix.hiesmayr@univie.ac.at}

\affiliation{
$^1$ Faculty of Physics, University of Vienna, Boltzmanngasse 5, 1090 Vienna, Austria\\
$^2$  Department of Theoretical Physics and Astrophysics, Masaryk University,
Kot\'a\v{r}\'ska 2, 61137 Brno, Czech Republic
}


\begin{abstract}
We show a general approach for detecting genuine multipartite entanglement (GME) and partial inseparability in many-body-systems by means of macroscopic observables (such as the energy) only. We show that the obtained criteria, the ``GME gap'' and ``the k-entanglement gap'', detect large areas of genuine multipartite entanglement and partial entanglement in typical many body states, which are not detected by other criteria. As genuine multipartite entanglement is a necessary property for several quantum information theoretic applications such as e.g. secret sharing or certain kinds of quantum computation, our methods can be used to select or design appropriate condensed matter systems.
\end{abstract}

\maketitle

\section{Introduction}
Quantum entanglement is undoubtedly one of the most prominent and unique features of quantum mechanics. It is not only essential for technological applications
 such as quantum computation (see e.g. \cite{compute}) or quantum cryptography (e.g. \cite{crypto}) and quantum secret sharing (e.g. \cite{qss1}), but also
exists in a wide range of non-artificial systems. While the actual role of entanglement in nature is still widely unclear and attracts much attention and
speculation (entanglement might e.g. be responsible for the high efficiency of phenomena such as photosynthesis \cite{photo1,photo2}, navigational orientation
of animals \cite{birds}, the imbalance of matter and antimatter in our universe~\cite{HiesmayrCP} or evolution itself \cite{evolution}), it is undoubted that entanglement at least exists in nature and appears to be of high importance in simple systems and processes, such as e.g. in particle physics \cite{particlephys} or in crystal lattices and phase transitions \cite{phasetrans}.

Many body systems have been studied particularly intensively in the context of entanglement in the past decade (see e.g. Refs.~\cite{mbe1,mbe2,mbe3,mbe4,mbe5,mbe6,mbe7,mbe8,mbe9} or, for an overview, Ref.~\cite{mbesummary}). However, none of these works addressed the issue of genuine multipartite entanglement (GME) and partial separability, that is the topic of this letter. In fact, to the authors' best knowledge the work in Ref.~\cite{mbgme} is the only one even mentioning the problem of partial separability in many body systems. However, while the approach presented in this work is, in principle, capable of detecting GME, this is neither done nor mentioned. Furthermore, since the method only yields non-optimal GME-witnesses (due to the estimations in the construction), it cannot offer a satisfactory detection power.

Due to the complete lack of research in this direction, the role of GME and partial entanglement in many body systems is still completely unknown (while especially the former is known to be of grave importance to applications like quantum secret sharing \cite{qss2} or certain kinds of quantum computers \cite{measurementqc}, as well as to fundamental tests of quantum mechanics \cite{foundations1,foundations2}).

In this letter, GME in many body systems consisting of lattices of interacting spin-$\frac{1}{2}$- and spin-$1$-particles is investigated, using existing tools for detecting the different kinds of entanglement as well as approaches inspired by the studies of bipartite entanglement in many body systems. The latter can be generalised, such that macroscopic thermodynamical observables, such as energy or entropy of the composite system, act as multipartite witnesses for different kinds of partial and genuine multipartite entanglement.


This work is organised as follows. First, a few basic definitions are reviewed, such that then methods for detecting partial entanglement and GME in many body systems with macroscopic observables - the main result of this paper - can be formulated. Finally, the results will be discussed and illustrated in representative examples.

\section{Partial Separability \& Genuine Multipartite Entanglement}
A pure state $\ket{\Psi^k}$ is called $k$-separable, iff it can be written as a tensor product of $k$ factors $\ket{\psi_i}$, each of which describes one or several subsystems:
\beq \ket{\Psi} = \bigotimes_{i=1}^k \ket{\psi_i} \; .\eeq
A mixed state $\rho$ is called $k$-separable, iff it can be decomposed into a mixture of $k$-separable pure states:
\beq \rho = \sum_i p_i \ket{\Psi_i^k}\bra{\Psi_i^k} \eeq
where all $\ket{\Psi_i^k}$ are $k$-separable (possibly w.r.t. different $k$-partitions) and the $p_i$ form a probability distribution.\\
An $n$-partite state (pure or mixed) is called fully separable iff it is $n$-separable. It is called genuinely multipartite entangled (GME) iff it is not biseparable (2-separable). If neither of these is the case, the state is called partially entangled or partially separable.\\

\section{Detection Criteria}
 The probably most important macroscopic observables of many body systems are Hamiltonians, corresponding to energy measurements. Therefore, using Hamiltonians for entanglement detection is in principle a very promising approach, as used in the concept of the entanglement gap (as defined in Ref.~\cite{mbe3}), which we generalise with respect to GME and partial entanglement and will denote in similar fashion ``\textit{the GME gap}'' and ``\textit{the k-entanglement-gap}'' (respectively). It works as follows.

Consider a system whose dynamics are described by a Hamiltonian $\mathcal{H}$. As a consequence of the compactness of the set of $k$-separable states, there always has to be a unique and well-defined energy $E_{k-sep}$ such that
\beq E_{k-sep} = \min_{\psi \in \mathcal{S}_k} \bra{\psi}\mathcal{H}\ket{\psi} \eeq
where $\mathcal{S}_k$ is the set of all $k$-separable states. Note that due to the convexity of $\mathcal{S}_k$, it suffices to optimise over pure states in order to obtain an optimum over the whole set, which is computationally feasible (and thus useful) for low particle numbers (as will be explicitly illustrated in examples later).

If now the ground state of $\mathcal{H}$ is not $k$-separable (or, in the case of a degenerate ground state, if there is no $k$-separable state in the ground state manifold), this energy is bound to satisfy
\beq E_{k-sep} > E_0, \eeq
where $E_0$ is the ground state energy of $\mathcal{H}$. Therefore, any state $\rho$ satisfying
\beq E_\rho = \tr(\rho \mathcal{H}) < E_{k-sep} \eeq
is necessarily $k$-inseparable (or, in particular, GME if $k=2$). For bipartite systems, the energy interval between $E_0$ and $E_{2-sep}$ is called the entanglement gap \cite{mbe3}. In analogy to this, we call the energy interval between $E_0$ and $E_{k-sep}$ the $k$-entanglement-gap (or, for $k=2$, the GME-gap). Note that since the sets of $k$-separable states for different $k$ are convex subsets of the sets of $(k-1)$-separable states, these energy gaps necessarily satisfy
\beq E_0 \leq E_{2-sep} \leq E_{3-sep} \leq \cdots \leq E_{n-sep} \eeq
where $n$ is the number of subsystems.

From the quantum informational point of view, this is nothing else than using $\mathcal{H}$ as an optimal entanglement witness. Since any operator with an
entangled eigenvector corresponding to the minimal (or maximal) eigenvalue can be used to detect entanglement effectively by finding the minimal expectation value
 a separable state can attain and using this as the detection threshold. The same procedure can also be applied to the more sophisticated problem of GME-
 or $k$-inseparability-detection (by replacing separable states by biseparable or $k$-separable states, respectively).

While the $k$-entanglement-gap is probably the most straightforward and practical macroscopic entanglement witness for partial entanglement in many body states, a similar approach can be applied to other observables as well. Also, in finite-dimensional systems, not only minima but also maxima can be used to construct entanglement witnesses (although these are of much less interest, as the states investigated in many body systems are usually ground states or thermal states, which typically are much closer to the ground state than to the highest excited state).

Furthermore, other methods of detecting entanglement by means of macroscopic observables can also often be generalised to detect $k$-inseparability and GME. As an example, consider the method in Ref.~\cite{entropy}. The starting point is very similar to the one above: a function is minimised over all separable states, leaving all lower function values for states which can therefore be recognised to necessarily be entangled. However, in this case, the function is not a simple thermodynamical observable on a state, but the quantum relative entropy
\beq S(\rho | \sigma) = \tr(\rho \ln \rho - \rho \ln \sigma)\;. \eeq
By chosing $\rho = \ket{E_0}\bra{E_0}$ as the ground state of $\mathcal{H}$ and
\beq \sigma = \frac{1}{Z} \sum_{i} e^{-\frac{E_i}{kT}} \ket{E_i}\bra{E_i} \eeq
as the thermal state with arbitrary temperature $T$ (where $k$ is Boltzmann's constant, $Z$ is the partition function, and the $E_i$ are the eigenvalues corresponding to the eigenvectors $\ket{E_i}$ of the Hamiltonian), this entropy becomes equal to the von-Neumann-entropy of the thermal state $\sigma$. Now, if
\beq S(\sigma) = S(\ket{E_0}\bra{E_0}\ |\sigma) < \min_{\omega \in \mathcal{S}} S(\ket{E_0}\bra{E_0}\ |\omega)\;, \eeq
then $\sigma$ is detected to be entangled. Since the right hand side of the above inequality only depends on $\ket{E_0}$, knowledge of the ground state in principle suffices to use this criterion. By simply replacing separable states by $k$-separable states, one obtains a similar criterion for detecting $k$-inseparability in thermal states: If
\beq S(\sigma) = S(\ket{E_0}\bra{E_0}\ |\sigma) < \min_{\omega \in \mathcal{S}_k} S(\ket{E_0}\bra{E_0}\ |\omega)\;, \eeq
then $\sigma$ is detected not to be $k$-separable (or to be GME, in the case $k=2$).

However, as this criterion only works for thermal states and requires optimisation over all mixed separable or $k$-separable states (as the optimised function is not linear), it is much less useful for practical and computational reasons than the concept of the entanglement gap or the $k$-entanglement-gap, respectively.

\section{Examples}
\noindent\emph{Example 1:} Consider a lattice of spin-$\frac{1}{2}$-particles with nearest-neighbor-interaction, described by the commonly used Heisenberg model Hamiltonian \cite{mbesummary}
\begin{eqnarray} \mathcal{H} &=&  \frac{J}{2} \sum_{\left\langle i,j\right\rangle} \left[(1+\gamma) \sigma^x_i\sigma^x_j + (1-\gamma) \sigma^y_i\sigma^y_j\right.\nonumber\\
&&\left. + 2\Delta \sigma^z_i\sigma^z_j\right] - h \sum_i \sigma^z_i \;, \end{eqnarray}
where the sum over $\left\langle i,j\right\rangle$ runs over all pairs of nearest neighboring particles. Note that this Hamiltonian can describe a one-, two- or three-dimensional lattice with arbitrary ordering structure, depending on the choice of index pairs $\left\langle i,j\right\rangle$ in the sum (describing the elementary cell of the lattice).

\begin{figure}[ht!]\centering\includegraphics[width=8cm]{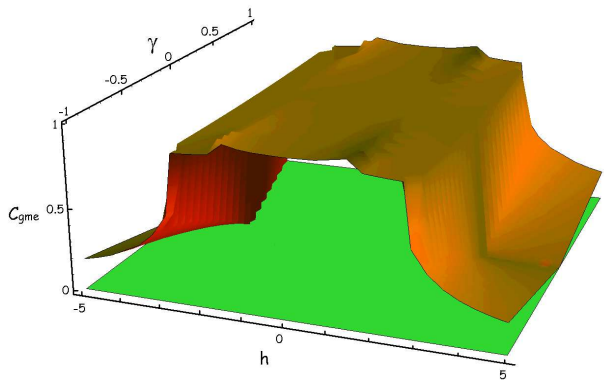}\caption{Illustration of the GME content of the ground state of $\mathcal{H}$ with a two-by-two particle quadratic lattice, for different values of $h$ and $\gamma$, measured by the gme-concurrence \cite{gmeconc}. It can be seen that the amount of GME only slightly depends on $\gamma$ and is near its maximally possible value of 1 for $|h|<2$, while decreasing monotonously with increasing $|h|$ (approaching zero asymptotically).}\label{fig_cgme}\end{figure}

For this example, let us choose the cubic antiferromagnetic case $J > 0$ with the parameter values $\gamma=[-1,1]$ and $\Delta = 1$. W.l.o.g. we can even choose $J = 1$ and measure all energies in units of $J$). This Hamiltonian's ground state is not only GME for large intervals of the external magnetic field $h$, but is also very strongly GME, as can be measured e.g. by the gme-concurrence $C_{gme}$ \cite{gmeconc} (a quantity which is zero for biseparable states and one for maximally GME states and can be derived analytically for pure states), as shown in Fig.~\ref{fig_cgme}. Thus, it is an optimal testing field for our criterion.

In order to test its detection power for mixed states as well, we compare it to some of the strongest criteria for GME known so far: A set of nonlinear inequalities of density matrix elements \cite{hmgh,dicke}, which we denote by $Q_i$ ($0 \leq i \leq \frac{n}{2}$):

\beq Q_0 &=& \left|\bra{0}^{\otimes n}\rho\ket{1}^{\otimes n}\right|\nonumber\\
&&- \sum_{\gamma} \sqrt{ \bra{0}^{\otimes n}\otimes\bra{1}^{\otimes n} \mathcal{P}_{\gamma_1} \rho^{\otimes 2} \mathcal{P}_{\gamma_1}^\dagger \ket{0}^{\otimes n}\otimes\ket{1}^{\otimes n}}\nonumber\\ \eeq
where the sum runs over all possible bipartitions $\gamma$ of the $n$-partite system and the permutation operator $\mathcal{P}_{\gamma_1}$ swaps all subsystems belonging to the first part of $\gamma$ between the two copies of $\rho$, and
\beq Q_m &=& \sum_{\{\alpha,\beta\}} \left(\left|\bra{d_\alpha}\rho\ket{d_\beta}\right|\right.\nonumber\\
 &&\left.- \sqrt{ \bra{d_\alpha}\otimes\bra{d_\beta} \mathcal{P_\alpha} \rho^{\otimes 2} \mathcal{P}_\alpha^\dagger \ket{d_\alpha}\otimes\ket{d_\beta}}\right) \nonumber\\
 &&- m(n-m-1) \sum_{\alpha}\bra{d_\alpha}\rho\ket{d_\alpha} \eeq
for $1\leq m \leq \frac{n}{2}$, where the first sum runs over all subsets $\alpha$ and $\beta$ of $\{1,2,3,...,n\}$ which satisfy $|\alpha|=|\beta| = m$ and $|\alpha \cap \beta| = m-1$, the permutation operator $\mathcal{P}_\alpha$ is the same is in the previous expression (swapping all subsystems contained in $\alpha$) and $|d_\alpha\rangle=\bigotimes_{i \not\in \{\alpha\}} |0\rangle_i\bigotimes_{i \in \{\alpha\}} |1\rangle_i$.

If any $Q_i > 0$, the state under investigation is detected to be GME (while a value lower than or equal to zero does not imply any statement whatsoever about GME, i.e. the criteria are sufficient but not necessary for GME). While each $Q_i$ is sensitive to a specific kind of GME (and thus capable of revealing different kinds of information on investigated states), a combination of all these criteria yields a well-distributed detection power.

The mixed states of highest interest in many body physics are thermal states, given by the Boltzmann distribution
\beq
\rho = \frac{1}{Z} \sum_i e^{-\frac{E_i}{kT}} \ket{E_i}\bra{E_i}\;, \eeq
where $E_i$ is the $i$-th energy eigenvalue, corresponding to the eigenvector $\ket{E_i}$ of $\mathcal{H}$, $Z$ is a normalisation factor (called the partition function) and $kT$ is the temperature multiplied by Boltzmann's constant. While it is clear from Fig.~\ref{fig_cgme}, that this state is GME for values of $kT$ close to zero, the important questions are: How far can $kT$ rise without $\rho$ becoming biseparable? What regions of $k$-inseparability can be detected for $k>2$?

\begin{figure}[ht!]\centering\includegraphics[width=7cm]{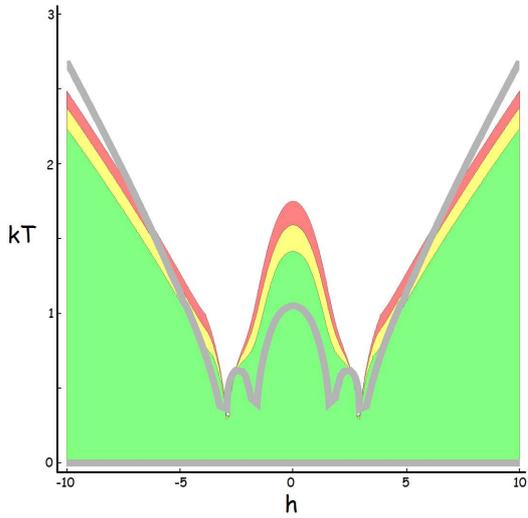}\caption{Illustration of the detection range of our criterion for the thermal states of the discussed Hamiltonian $\mathcal{H}$ (green, yellow and red for $k=2$, $k=3$ and $k=4$, respectively) compared to the GME detection inequalities $Q_i$ (Refs.~\cite{hmgh,dicke}) combined (area between the two gray lines). It can be seen, that detection via the GME gap is particularly strong in areas with highly GME ground states (i.e. near $h=0$), while it is slightly weaker (by comparison) in weakly GME ground state areas. As long as the ground state is GME, large intervals of $kT$ are detected to be so as well. Note, that the energy gap between the ground state and the first excited state is usually rather big (see Ref.~\cite{mbe2}), which leads to particularly high temperature GME for high magnetic field magnitudes (as an increasing temperature in this case only comparatively slowly leads to increasingly mixed states).}\label{fig_thermal}\end{figure}

In Fig.~\ref{fig_thermal}, the detection power of our criterion is compared to the combined detection inequalities $Q_i$ for a two-by-two particle (i.e. two-dimensional quadratic) lattice for the choice $\gamma=1$.

\begin{figure}[ht!]\centering\includegraphics[width=8cm]{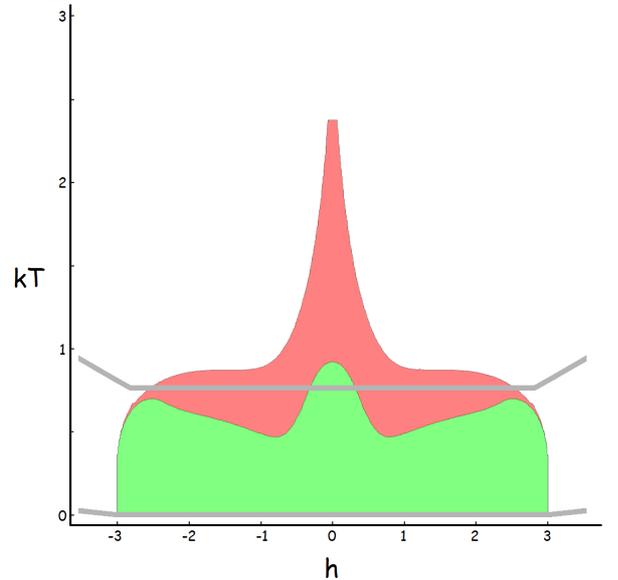}\caption{Illustration of the detection power of the GME gap witness (green) and the 3-entanglement-gap witness (red) for the discussed spin-1-Hamiltonian, compared to the joint detection power of the criteria $Q_i$ (area between the two gray lines). While by construction the $k$-entanglement-gap witnesses do not detect any entanglement for $|h|>3$ (since the corresponding ground states are separable), for $|h|<3$ the witnesses detect significantly large areas of states to be entangled / GME.}\label{fig_qutrits}\end{figure}

\noindent\emph{Example 2:} In order to illustrate that the concept of the $k$-entanglement- and GME-gap can also be applied to less common Hamiltonians and higher spin systems, consider a chain of $n$ spin-$1$-particles with nearest-neighbor coupling, described by the Hamiltonian
\beq \mathcal{H} = \sum_{i=1}^n \left(\vec{S}_i\cdot\vec{S}_{i+1} + \beta (\vec{S}_i\cdot\vec{S}_{i+1})^2\right) + h \sum_{i=1}^n S_i^z\;, \eeq
where $\vec{S}_i = (S_i^x,S_i^y,S_i^z)$ is the vector of spin-1-operators. Let us choose $n=3$ and $\beta = 1$ (the choice of $\beta$ only slightly influences the separability-behaviour of the Hamiltonian's thermal states and thus does not play a significant qualitative role in our considerations). For these settings, the ground state is GME for $|h| < 3$. It can be seen from Fig.~\ref{fig_qutrits} that in this case also significant areas of entanglement and GME are detected (again, in some areas better than with the previously known criteria $Q_i$, in other areas worse).

\section{Conclusion}
The introduced concept of the $k$-entanglement-gap - and, in particular, the GME (genuine multipartite entanglement) gap - i.e. the gap between the energy value minimised over all $k$-separable states and the ground state energy, offers a simple detection criterion for partial entanglement and GME in a wide variety of systems, the only requirement being that the ground states are not separable and the Hamiltonian is known. Since its expectation values for $k$-separable states are bounded, any expectation value exceeding these bounds has to be due to a $k$-inseparable (or, for $k=2$, GME) state. The detection power of this tool compares to some of the strongest developed criteria known so far for GME detection. It can be measured simply by means of a single macroscopic observable, the energy. Since $k$-inseparability detection (like bipartite entanglement detection) is in general an NP-hard problem, no computable necessary and sufficient criteria exist, leaving sufficient-only criteria the only way to tackle this problem. GME detection in many body systems is a crucial step not only in verifying experimental advances in this field, and also may eventually allow for these systems to be utilised in quantum informational technology.  Since in experimental scenarios, the energy is often rather difficult to measure directly, related quantities - such as e.g. the heat capacity \cite{heatcap} - can be measured instead.

We also showed that the concept can be applied to other thermodynamical quantities as well, such as the relative entropy which is a powerful tool in analysing different quantum information theoretic issues. Furthermore, the concept can be generalised in order to improve its detection range further - for example, a similar quantity to the $k$-entanglement-gap could be defined by maximising (instead of minimising), thus detecting partial entanglement in the upper end of the energy spectrum - and to be sensitive to different types of multipartite entangled states.

As different kinds of multipartite entanglement (especially GME) are of grave importance not only to several quantum informational applications, but also to fundamental tests of quantum mechanics, our criterion can be used to experimentally verify the presence of such states and thus aid in investigating these foundational issues.\\
\\
\textbf{Acknowledgements:}\\
We thank Marcus Huber for very fruitful discussions. The project is funded from the SoMoPro programme. Research of BCH leading to these results has received a financial contribution from the European Community within the Seventh Framework Programme (FP/2007-2013) under Grant Agreement No. 229603. The research is also co-financed by the South Moravian Region. AG gratefully acknowledges the Austrian Research Fund project FWF-P21947N16.

\end{document}